\documentclass[11pt]{article}
\usepackage{amsmath}
\usepackage[dvips]{graphicx}
\usepackage{amsfonts}
\usepackage[english]{babel}
\newcommand{\be}{\begin{equation}}
\newcommand{\ee}{\end{equation}}
\newcommand{\bea}{\begin{eqnarray}}
\newcommand{\eea}{\end{eqnarray}}
\newcommand{\f}[2]{\frac{#1}{#2}}

\title{Problems with Tunneling of Thin Shells from Black Holes}
\author{Borun D. Chowdhury}

\begin{document}

\begin{center}
{\LARGE  Problems with Tunneling of Thin Shells from Black Holes}
\\
\vspace{18mm}
{ \bf Borun D. Chowdhury\footnote{borundev@pacific.mps.ohio-state.edu}}\\

\vspace{8mm}
Department of Physics,\\ The Ohio State University,\\ Columbus,
OH 43210, USA\\
\vspace{4mm}
\end{center}
\begin{abstract}
It is shown that $exp(-2\; Im(\int p \; dr))$ is not invariant under canonical transformations in general. Specifically for shells tunneling out of  black holes, this quantity is not invariant under canonical transformations.  It can be interpreted as the transmission coefficient only in the cases in which it is invariant under canonical transformations. Although such cases include alpha decay, they do not include the tunneling of shells from black holes. The simplest extension to this formula which is invariant under canonical transformations is proposed. However it is shown that this gives half the correct temperature for black holes.
\end{abstract}
\newpage

\section{Introduction}

Black holes were shown to radiate thermally by Hawking \cite{Hawking:1974sw}. This was a result of semi-classical gravity in which field theories are quantized on classical curved spacetime backgrounds. It was also suggested by Hawking and Hartle \cite{Hartle:1976tp} that Hawking radiation could be modeled as  tunneling of particles across the horizon of the black hole. Hawking radiation was calculated for the emission of test particles (not affecting the background).

Hawking radiation poses the so called Information Paradox. Two completely different systems can collapse to form identical black holes and evaporate away leaving behind completely identical thermal radiation. In this way the information of the original collapsing systems is completely lost. For further details one can refer to \cite{Giddings:1995gd}.

One of the approaches taken to try to fix the paradox was to include the self-gravitational correction to the radiation. It was hoped that if the emitted particle's effect on spacetime curvature was also taken into account the radiation would not be thermal and the paradox would be resolved.

Israel had derived the equations of motion of self-gravitating shells \cite{Israel:1966rt} a decade before Hawking's derivation of Hawking radiation. The results were, however, derived from considering the Einstein's equations and not from an action. Without an action it is not clear how to quantize the shells.

An action for the self-gravitating shell was proposed by Kraus and Wilczek \cite{Kraus:1994by}. Parikh and Wilczek \cite{Parikh:1999mf} worked on the idea of tunneling of shells using the action. They computed the quantity $exp(-2 Im (\int p \; dr))$ where the integration domain includes the horizon (which makes the action imaginary due to a pole in the momentum). This quantity was then taken to be equal to the transmission coefficient as is done for alpha particle emission. This method was then applied to several different black holes in various dimensions by many authors \cite{Akhmedov:2006un},\cite{Wu:2006nj},\cite{Hu:2006ek},\cite{Arzano:2005jt},\cite{Radinschi:2005ap},\cite{Nadalini:2005xp},\cite{Medved:2005yf},\cite{Kerner:2006vu}.

The importance of \cite{Parikh:1999mf} seemed to be that it offered a correction to Hawking radiation making it non thermal. Non thermality of the radiation was taken as a possible sign of resolution of the black hole information paradox \cite{Melnikov:2002qd},\cite{Zhang:2005mn},\cite{Einhorn:2005bi}. It was also proposed that the non-thermality of the radiation had an effect on the inflationary vacuum \cite{Greene:2005wk}.

Tunneling is the system penetrating a barrier quantum mechanically which would have been impossible classically. Although there is nothing wrong with assuming that the same physics would be at work in curved space time it would certainly be worth investigating if the formula  $exp(-2 Im (\int p \; dr))$ could be used for black hole emission. It turns out that if the particles going either ways face the same barrier then the said formula is the same as  $exp(- Im (\oint p \; dr))$. This guarantees that the transmission coefficient is invariant under canonical transformations. However such is not the  case for black holes as infalling shells can fall through classically while outgoing shells cannot come out. If we were to assume that the generalized formula is the latter one then using  $exp(-2 Im (\int p \; dr))$ gives incorrect results.

In this paper we work with the action for the shell due to Gladush \cite{Gladush:2000rs}. Gladush was able to reproduce Israel's junction conditions \cite{Israel:1966rt} from the action thus lending credibility to it. The action also gives Israel's equations of motion for the self gravitating shell \cite{Israel:1966rt} on variation. Additionally the paper gives two equivalent actions for the shell one for the outside manifold and one for inside.  We calculate  $ \int p \; dr$ from these two actions  and show that the results are different from each other and from the result of \cite{Parikh:1999mf}. The first point is not really a problem since we do not expect to get a correct tunneling probability from the inside manifold anyways since the horizon in the inside manifold is at a different radius than the one for asymptotic observers.

We also observe that the expression for transmission coefficient which is invariant under canonical transformations namely $exp(- Im \oint pdr)$ gives half the temperature as that of the black hole showing that all of these simple extensions of flat spacetime tunneling to black holes do not give reasonable answers and further work is needed to understand curved spacetime tunneling.
\newline
\newline
{\bf Outline of the paper} 

\begin{itemize}
\item In section \ref{ch:CanTrans} we review tunneling and see a simple extension which is invariant under canonical transformations.
\item We go over the derivation of the conventional tunneling model from \cite{Parikh:1999mf} in section \ref{ch: Parikh's Tunneling Model}.
\item In section \ref{ch:Preliminaries} we explain the geometry of the problem. 
\item We summarize the equations of motion of massless and massive shells from Israel's method \cite{Israel:1966rt} in section \ref{ch:MtnThShell}.
\item In section \ref{ch: Reduced Gravity Action For a Thin Shell} we derive the same equations of motion by varying Gladush's action \cite{Gladush:2000rs}.
\item We calculate the different possible expressions for the transmission coefficient in section \ref{ch:Tunneling} and show how the answer that is invariant under canonical transformations gives a temperature half that of the correct one. Thus we conclude that more work is needed to get an expression for tunneling in black hole backgrounds.
\item In \ref{ch:Israel's Junction Conditions} we derive the equation of motion by Israel's method.
\item In \ref{ch:Eddington Finkelstein Coordinates and The Equation of Motion} we explain the infalling Eddington Finkelstein coordinates.
\item In \ref{altdrvfrmactn} alternate derivations of  $exp(- 2 Im(\int pdr))$ are given.
\item In \ref{ch:Meeting Newton's Laws} we motivate the action from the equations of motion.
\end{itemize}

\section{Tunneling}  \label{ch:CanTrans} 

Tunneling is  said to happen when the initial and final states are separated by a barrier which cannot be classically crossed because the system does not have enough energy. Tunneling is a quantum mechanical phenomenon which has been well understood in flat spacetime. However quantum mechanics in flat spacetime is unitary and it is not so, in general, in curved spacetime. Although it seems reasonable that tunneling will continue to happen in curved spacetime it is not clear immediately how to generalize the formulae. In this section we first go over tunneling in unitary quantum mechanics along the lines of \cite{Rubakov:2002fi}. We show that there are two expressions for the tunneling transmission coefficient which are equivalent when the barrier is not sensitive to the direction of motion. We then proceed to show that in case the barrier is sensitive to direction of motion, invariance under canonical transformation allows only one of the expressions to be consistent.

\subsection{Review of Tunneling} 

In semi classical approximation the amplitude to go from an initial state to a final state is given by
\be
\langle x_f| x_i \rangle = e^{iS_{stationary}}
\ee
where the action is evaluated at the stationary point. 

\begin{figure}[htbp] 
   \centering
   \includegraphics[width=3in]{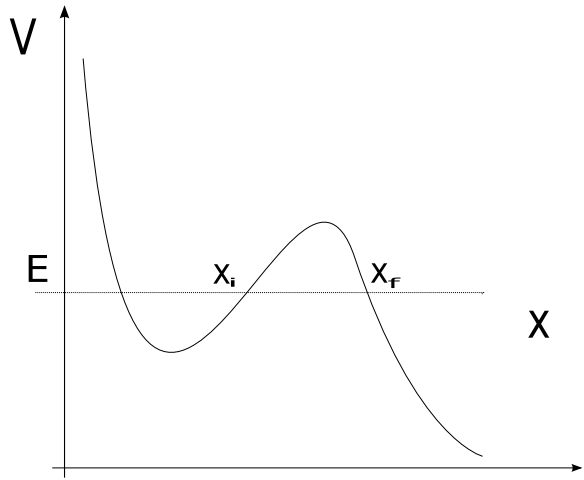} 
   \caption{Regular tunneling problem}
   \label{fig:tunneling}
\end{figure}

Let us now evaluate this for a tunneling problem. The action is given as

\be
S=\int \left[ \f{m}{2} \left(\f{d x}{dt}\right)^2 - V(x) \right] dt
\ee
and the first integral of motion is

\be
\f{m}{2} \left(\f{d x}{dt}\right)^2+ V(x)=E
\ee
and the action needs to be evaluated along a path given by this equation. This equation does  not have a solution in real time in the region where $V(x) > E$ or in other words between the points $x_i$ and $x_f$ in the figure \ref{fig:tunneling}. However if we do a formal substitution $t= - i \tau$ to get the action

\be
S = i \int \left[\f{m}{2} \left( \f{dx}{d \tau}\right)^2 + V(x) \right] d\tau = i S_E
\ee

The first integral is written in terms of this imaginary time as

\be
E= \f{m}{2} \left(\f{dx}{dt}\right)^2 + V(x) =  - \f{m}{2} \left(\f{dx}{d \tau}\right)^2 + V(x) 
\ee
which has a solution in the region classically forbidden. The action evaluated in the classically forbidden region is

\bea
S &=& i \int_0^{\tau_0} (2 V - E) d \tau \nonumber \\
&=& i \int_0^{\tau_0} 2 (V-E)  d\tau  + i E \tau_0 \nonumber \\
&=& i \int_{x_i}^{x_f} 2(V-E)  \sqrt{\f{m}{2(V-E)}} dx + i E \tau_0
\eea
where $\tau_0$  is the Euclidean time taken for the system to go from $x_i$ to $x_f$  in the potential $-V(x)$ with an energy $-E$ in figure  \ref{fig:tunneling}. So we have

\be
\langle x_f| x_i \rangle = e^{- \int_{x_i}^{x_f} \sqrt{2 m (V-E)}  dx} e^{- E \tau_0}
\ee
Since the potential is independent of weather the particle is going from $x_i$ to $x_f$ or in the opposite direction we have

\be
\langle x_i | x_f \rangle = e^{- \int_{x_f}^{x_i} \sqrt{2 m (V-E)}  dx} e^{- E \tau_0}
\ee
and the transmission coefficient to go from $x_i$ to $x_f$ (or equivalently from $x_f$ to $x_i$) is 

\be
\Gamma= \langle x_i | x_f \rangle \langle x_f| x_i \rangle = | \langle x_f| x_i \rangle |^2 =  e^{- \oint  \sqrt{2 m (V-E)}  dx} e^{-2 E \tau_0}
\ee
Where replacing the Euclidean time period $2\tau_0$ by the inverse temperature $\beta$ we have the transmission coefficient of a particle of energy $E$ tunneling through a barrier at inverse temperature $\beta$ as

\be
\Gamma= e^{-\beta E} e^{- \oint  \sqrt{2 m (V-E)}  dx} 
\ee
This formula is usually written in a different form. With the realization that momentum is given by

\be
p^{2} = 2m(E-V)
\ee
and the independence of the potential on the direction of motion we have the transmission coefficient 

\be
\Gamma= e^{-\beta E}  e^{-2 Im(\int p dx)}  =  e^{-\beta E}  e^{- Im(\oint p dx)}
\ee

\subsection{Tunneling Calculations and Canonical Transformations} 
The action evaluated over an open path in phase space is in general different in different in canonically equivalent frames

\begin{equation}
S=S' + \int_{initial}^{final} dF = S' + F_{final} - F_{initial}
\end{equation}
where F is the generating functional for the canonical transformation and $S$ and $S'$ are the actions along the phase space path in the two canonically equivalent frames.
\newline
\newline
For an open path in phase space we have, in general, in two canonically equivalent frames

\begin{equation}
\int p dx \ne  \int P dX
\end{equation}
If we take the transmission coefficient of tunneling as

\be
\Gamma= e^{-2 Im \int p dx}
\ee
in one canonical frame then in general the answer will be different in a different canonical frame. The reason we do not have this problem in the usual examples of tunneling phenomenon is that tunneling in both directions is equally suppressed. Being a transmission coefficient the answer cannot depend on the canonical frame chosen so if the barrier is sensitive to direction then the tunneling transmission coefficient that should be used is
\be
\Gamma= e^{- Im \oint p dx}
\ee

\subsection{Tunneling from black hole horizons}

We saw that we need to understand tunneling better in curved spacetime. We found that the more commonly used expression 

\be
e^{-2 Im \int p dx}
\ee
cannot be used if the barrier is sensitive to direction of motion. Infalling shells face no barrier at all and for them

\be
\langle in | out \rangle = e^{i \int_{horizon+}^{horizon-} p dr} =1
\ee
 where we have integrated from just outside the horizon to just inside. Outgoing shells however cannot cross the horizon.  As was show in \cite{Parikh:1999mf}-\cite{Kerner:2006vu} and will be shown in detail later in this paper, for them $\int_{horizon} p dr$ has an imaginary part from a pole. This gives
 
\be
\langle out | in \rangle = e^{-Im \int_{horizon+}^{horizon-} p dr}  \ne 1
\ee
 where the exact value depends on the expression for the momentum and will be given later in the paper. So we see that
 
 \be
 \langle in | out \rangle \ne  \langle out | in \rangle^*
 \ee
 and so we cannot use
\be
e^{-2 Im \int p dx}
\ee 
as the tunneling transmission coefficient but must instead use

\be
P= e^{- Im \oint p dx} = e^{-Im \int_{hor+}^{hor-} p dr} e^{-Im \int_{hor-}^{hor+} p dr} 
\ee
which on account of infalling shells facing no barrier becomes

\be
P=e^{-Im \int_{hor-}^{hor+} p dr} 
\ee
Thus the results of tunneling probabilities in \cite{Parikh:1999mf}-\cite{Kerner:2006vu} are squares of what they should be. However it was pointed out in these papers that the answers they got gave Hawking temperature for test shells. If one were to take the square root of their answers then one would get the temperature off by a factor of two.

\section{Conventional Tunneling Model: A Review} \label{ch: Parikh's Tunneling Model}

In this section the tunneling model in \cite{Parikh:1999mf}, which we refer to as the conventional model for tunneling, will be reviewed for completeness. That model used the action for shells found in \cite{Kraus:1994by}. Only null shells were considered. The Hamiltonian of the shell gravity system according to \cite{Kraus:1994by} and \cite{Parikh:1999mf} is the ADM mass. For massless outgoing shells in Eddington Finkelstein coordinates we get the equation of motion
\begin{eqnarray}
\frac{dr}{dt}=\frac{1-\frac{2M}{r}}{2} \label{ParikhNullSpeed}
\end{eqnarray}
The imaginary part of  $\int p dr$ was calculated by the relation
\begin{eqnarray}
\int p dr = \int \int dp dr
\end{eqnarray}
From the relation $\dot{r}=\frac{dH}{dp}$ formally the integral was rewritten as
\begin{eqnarray}
\int p dr = \int \int \frac{dH}{\dot{r}}dr
\end{eqnarray}
The Hamiltonian was taken to vary from $M$ to $M-\omega$ where $\omega$ was the shell's energy. With this and (\ref{ParikhNullSpeed}) the value for imaginary part of $\int pdr$ was found out by integrating over the horizon and going under the pole as
\begin{eqnarray}
Im(\int_{horizon} p dr) &=& Im( \int_{horizon} \int_{M}^{M-\omega} \frac{ dH dr}{\frac{dr}{dt}} )\nonumber \\
&=& Im( \int_{horizon} \int_{M}^{M-\omega} \frac{2 r dH dr}{r- 2H} )\nonumber \\
&=& \pi \int_{M}^{M-\omega} 4H dH \nonumber \\
&=& \pi 4 \omega (M- \frac{\omega}{2})
\end{eqnarray}
The sign comes out to be positive if $r_{in} > r_{out}$ which was explained by saying that the horizon shrinks while emitting the shell so the tunneling process starts from just behind the horizon to emerge just outside the shrunken horizon.

Thus the tunneling transmission coefficient was found to be
\begin{eqnarray}
\Gamma = e^{-8\pi \omega M (1-\frac{\omega}{2M})} \label{parikhprobability}
\end{eqnarray}

\section{Geometry and Causality} \label{ch:Preliminaries}

\begin{figure}[htbp] 
   \centering
   \includegraphics[width=2in]{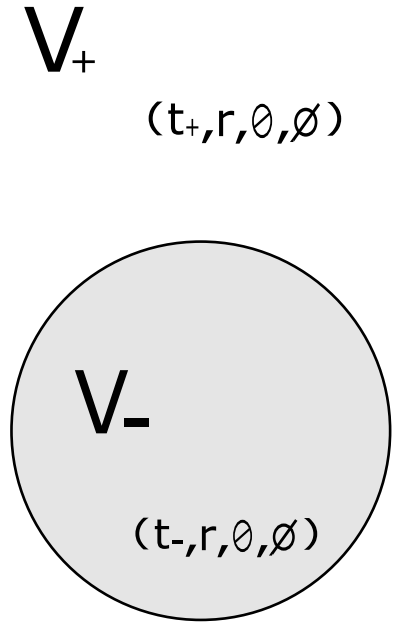} 
   \caption{The geometry around a singular spherical hypersurface}
   \label{fig:hypersur-geom}
\end{figure}

{\bf Geometry}
\newline
\newline
The geometry of space time with a thin shell is non trivial and Israel \cite{Israel:1966rt} showed that \emph{a singular hypersurface divides the space time that it moves in into two regions which do not share the same mass. Although it is possible to have coordinate charts which are continuous across the hypersurface for non static coordinates, it is not possible to have continuous coordinates across the hypersurface for static coordinates.} In our case the region inside/outside the shell will be referred to as $V_{\pm}$ and their mass parameters as $M_{\pm}$. The geometry is shown in fig. \ref{fig:hypersur-geom}.  \emph{The horizons for both the regions are at different values of the radial parameter}.  

Classically the motion of the shell can be specified completely by saying it is at $r=R(\tau)$ at $t_-=T_-(\tau)$ in terms of internal coordinates where $\tau$ is the proper time of the shell. It can also be specified completely by saying it is at $r=R(\tau)$ at $t_+=T_+(\tau)$ in terms of external coordinates. 

The fact that the angular variables can be taken to be the same on both sides of the shell is because the geometry of the shell on both side is $S^2$. The radial coordinate can be taken to be the same for the same reason as it is defined as $4\pi r^2\equiv A$ where A is the area of the $S^2$. The continuity of coordinates across the shell, however, cannot be maintained for the time coordinate. The presence of the shell causes the mass parameters to be different on both sides. Considering the motion of a null shell shows that \emph{the time coordinates have to be different on both sides for static coordinates}. 

In Schwarzschild coordinates the metric on both sides of the shell is 

\begin{eqnarray}
ds_{\pm}^2&=&-f_{\pm}dt_{\pm}^2+f_{\pm}^{-1}dr^2 + r^2d\Omega^2 \nonumber \\
f_{\pm}&=&1-\frac{2M_{\pm}}{r} \label{lnelement}
\end{eqnarray}
When the shell is at a radial coordinate R, the metric on the shell, due to spherical symmetry can be taken as

\begin{equation}
ds_{\Sigma}^2\equiv -d\tau^2 + R^2 d\Omega^2
\end{equation}
which defines the proper time on the shell. The relationship of the shell's proper time with the manifold coordinates is explained and equations of motion worked out in the \ref{ch:Israel's Junction Conditions}. 

Since we have used up $\pm$ for distinguishing which manifold is being discussed we will use the symbol $\oslash$ to signify $+$ for outgoing shells and the same symbol to signify $-$ for infalling shells whenever such distinction is required (specifically for quantities calculated in Eddington-Finkelstein coordinates).
\newline
\newline
{\bf Causality}
\newline
\newline
Although the shell divides spacetime into two manifolds each with a horizon, the two horizons here should not to be confused with two horizons of a two charge solution like a Reissner Nordstrom black hole. 

\begin{figure}[htbp] 
   \centering
   \includegraphics[width=5in]{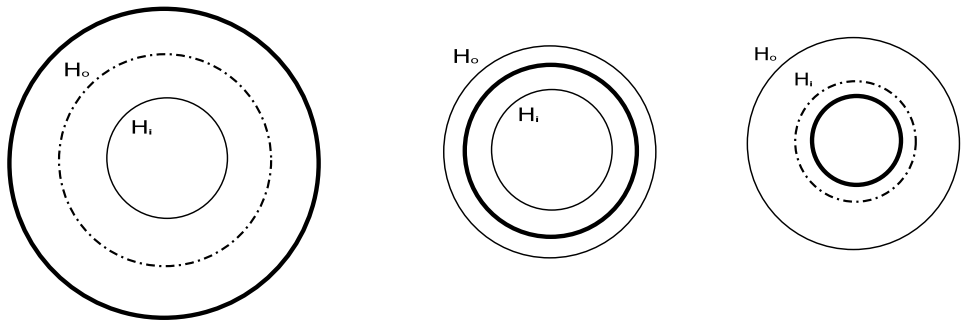} 
   \caption{The dark circle shows the shell at various values of the radius parameter. The light circles show the position of the horizon. Inside the shell is $V_-$ and outside $V_+$. Depending on the position and future of the shell the horizons in $V_\pm$ may or may not exist.}
   \label{fig:causality}
\end{figure}

We will refer to the horizon of the outer manifold as $H_o$ and that of the inner manifold as $H_i$. Due to positivity of mass, $H_o$ will always be at a larger radial parameter than $H_i$. We could have three scenarios as shown in fig. \ref{fig:causality}. The dark circle is the shell and the dotted circle is where the outer or inner horizons, as the case may be, would have been had the appropriate manifold extended to that point. The light circles are the outer and inner horizons.

The shell could be outside $H_o$ as show in the first case. In this case the outer manifold, which reaches upto the shell, does not contain a horizon. All signals from the shell can reach the observer at infinity. The inner manifold however does have a horizon. Any signal emitted from inside $H_i$ cannot reach the shell (and hence cannot reach the observer at infinity either).

The second case is that the shell  is between  $H_i$ and $H_o$.  Then signals emitted from the shell cannot reach the observer at infinity. The signals emitted from inside $H_i$ cannot reach the shell. Any outgoing signal emitted from the region between $H_i$ and  the shell would eventually emerge into the outer manifold. It will however emerge inside $H_o$ and due to bending of the light cone will keep falling to lower values of the radial parameter. This argument also shows that once in the intermediate region the shell will eventually fall into $H_i$.

Finally, in the third case the shell is inside $H_i$. It will continue falling till it hits a singularity. Signals emitted from inside may or may not meet the shell depending on when the shell meets the singularity. If the signals meet the shell they will emerge in the outer manifold in an inward bent light cone and thus fall back to the singularity.

Thus classically once a shell crosses $H_o$ it will meet a singularity.

\section{Motion of a Thin Shell: Extending Israel's Solution} \label{ch:MtnThShell}

In this section the equations of motion of a thin shell moving in a Schwarzschild background are summarized. The details of the calculations can be found in \ref{ch:Israel's Junction Conditions}. 

\subsection{Motion of a Null Shell} \label{ch:Motion of a null shell}

A null shell will move along coordinates in such a way that the path length is zero. Thus the motion in outside and inside Schwarzschild coordinates from (\ref{lnelement}) is

\begin{eqnarray}
\big(\frac{dR}{dT_{\pm}}\big)^2=F_{\pm}^{2} \label{nullspeed} \equiv f_{\pm}^2(R)
\end{eqnarray}

There is the usual coordinate singularity at the horizon of Schwarzschild  coordinates (the shells seem to stop at the horizon) and we can go to some well behaved coordinate systems like Eddington-Finkelstein coordinates to study the motion through the horizon. That will be done in subsection \ref{ch:EFcood} for massive shells and can be generalized to the case of null shells.

\subsection{Motion of a Massive Shell} \label{ch:Motion of Massive Shell}

In this subsection we first summarize the equations of motion in Schwarzschild coordinates. However, we will be considering particles falling through the horizon so we then rewrite the equations of motion in infalling Eddington Finkelstein coordinates.

\subsubsection{Equations of Motion in Schwarzschild Coordinates}

The equations of motion are

\begin{eqnarray}
\dot{T}_{\pm}&=& \frac{\kappa_{\pm}}{F_{\pm}}  \nonumber \\
\dot{R}^2&=&\kappa_{\pm}^2 - F_{\pm} \nonumber  \\
\kappa_{\pm}&=&\frac{1}{m}\big(\Delta M \mp \frac{m^2}{2R}\big) \label{eqnofmtn}
\end{eqnarray}
where m is a constant of integration which can be interpreted as the rest mass of the shell and

\begin{eqnarray}
\Delta M &=& M_+ - M_- \nonumber \\
F_{\pm} &=&   1 - \frac{2M_{\pm}}{R}
\end{eqnarray}
We can eliminate the proper time on the shell and rewrite the equations of motion 

\begin{eqnarray}
\frac{dR}{dT_{\pm}} = \frac{F_{\pm} \sqrt{\kappa_{\pm}^2- F_{\pm}}}{\kappa_{\pm}} \label{speed}
\end{eqnarray}

It can be seen from comparing (\ref{nullspeed}) and (\ref{speed}) while using (\ref{eqnofmtn}) that the limit of a null shell can be obtained by taking the rest mass of shell to vanish, $m \to 0$.

We can compare this result with the known result for test particles. We know that for test particles the equations of motion are

\begin{eqnarray}
\dot{T}&=&\frac{\kappa}{F} \nonumber \\
\dot{R}^2&=& \kappa^2-F \nonumber \\
F&=&1- \frac{2M}{R}  \label{tsteqnofmtn}
\end{eqnarray}
For uncharged test particles $\kappa$ is a constant (and is one for test particles coming to a stop at $r=\infty$.)  Thus we see that if we drop the term of order $m^2$ from (\ref{eqnofmtn}) and take the particle to not influence the geometry (in other words drop the $\pm$s because we would not have two different inner and outer manifolds) we obtain  (\ref{tsteqnofmtn}).

We can now compare this general result with the more specific result of Israel's \cite{Israel:1966rt} self-gravitating shell. A self-gravitating  shell is given by  $M_-=0$ and $M_+=M$. This gives $\Delta M= M$ and the equations of motion reduce to

\begin{eqnarray}
\dot{T}_{\pm}&=& \frac{\frac{M}{m} \mp \frac{m}{2R}}{F_{\pm}}  \nonumber \\
\dot{R}^2&=&\big( \frac{M}{m} \mp \frac{ m}{2R}\big)^2 - F_{\pm} 
\end{eqnarray}
Israel specifically wrote the equation for $\dot{R}$ in terms of inner manifold quantities. In this case we have $F_-=1$ we get the equations of motion as

\begin{eqnarray}
\dot{T}_{\pm}&=& \frac{\frac{M}{m} \mp \frac{m}{2R}}{F_{\pm}}  \nonumber \\
\dot{R}^2&=&\big( \frac{M}{m} + \frac{ m}{2R}\big)^2 - 1
\end{eqnarray}
These are the same equations of motion as Israel's.

\subsubsection{Equations of Motion in Infalling Eddington Finkelstein Coordinates} \label{ch:EFcood}

To study objects falling into black holes we need to go to some coordinate map which covers the future horizon. In the case of a shell we need to do this for both the manifold inside and outside the shell. In this paper we choose Eddington Finkelstein coordinates. The coordinates and transformation laws are explained in the \ref{ch:Eddington Finkelstein Coordinates and The Equation of Motion} and here we summarize the equations of motion. As explained in section \ref{ch:Preliminaries} we use the symbol $\oslash$ to mean $+$ for outgoing shells and $-$ for ingoing shells.

\begin{eqnarray}
\dot{R}^2&=&\kappa_{\pm}^2-F_{\pm} \nonumber \\
\dot{T}_{\oslash,\pm}&=&\frac{\kappa_{\pm} \oslash \dot{R}}{F_{\pm}} \label{eqnofmtnEF}
\end{eqnarray}
Eliminating the proper time of the shell we get

\begin{eqnarray}
\acute{R}_{\oslash,\pm}&=& \frac{dR}{dT_{\pm}} \big |_{\oslash} = -(\kappa_{\pm}^2 - F_{\pm}) \oslash \kappa_{\pm}\sqrt{\kappa_{\pm}^2-F_{\pm}}  \label{eqnofmtnEFspeed}
\end{eqnarray}

We can see from (\ref{eqnofmtnEFspeed}) that the ingoing shell falls in through the horizon. The outgoing shell, however, never comes out of the horizon, which is the expected result.

\section{The Action of a Thin Shell} \label{ch: Reduced Gravity Action For a Thin Shell}

In this section the action for the thin shell system will be discussed and the equations of motion will be derived at by varying the same. This action was proposed by Gladush \cite{Gladush:2000rs}.

\subsection{The Action} \label{ch: The Action}

It was shown in \cite{Gladush:2000rs} that from the complete gravity action Israel's Junctions conditions can be derived. The complete gravity action gives the effective action for the shell when evaluated for the Schwarzschild solution in regions $V_\pm$. The effective action was shown to be

\begin{equation}
I^{\pm}_{sh}=-m \int \big( d\tau \mp U_{\alpha}dX^{\alpha}\big) \label{actioncovariant}
\end{equation}
where U is a gauge potential and for Schwarzschild coordinates $U=\{-\frac{m}{2R},0,0,0\}$ for a particular gauge choice.  The two actions ($\pm$) are for the same shell but in coordinates of inside or outside manifolds. It will be shown that both of them give the correct equation of motion and either should be sufficient to understand the complete motion classically.

\subsection{Equations of Motion by varying the Action} \label{ch: Varying the Action}

We vary the action to obtain the equations of motion first in Schwarzschild coordinates and then in infalling Eddington-Finkelstein coordinates.

\subsubsection{Equations of Motion in Schwarzschild Coordinates}

The action (\ref{actioncovariant}) in Schwarzschild coordinates is

\begin{eqnarray}
I^{\pm}_{sh}&=&-m \int \big( \sqrt{F_{\pm} - F^{-1}_{\pm} \acute{R}_{\pm}^2} \pm \frac{m}{2R}\big) dT_{\pm} \label{actionSC}
\end{eqnarray}
where $\acute{R}_{\pm} \equiv \frac{dR}{dT_{\pm}}$.	
Thus the Lagrangian is

\begin{equation}
L_{sh}^{\pm}=-m \sqrt{F_{\pm} - F^{-1}_{\pm} \acute{R}^2_{\pm}} \mp \frac{m^2}{2R} \label{lagrangianSC}
\end{equation}
The Lagrangian is independent of the coordinate time so the Hamiltonian is a constant of motion. The conjugate momentum of the shell is

\begin{eqnarray}
p_{sh}^{\pm} &=&\frac{\partial L_{sh}^{\pm}}{\partial \acute{R}_{\pm}} \nonumber\\
&=& \frac{m F_{\pm}^{-1} \acute{R}_{\pm}}{\sqrt{F_{\pm} - F^{-1}_{\pm} \acute{R}^2_{\pm}}}   \label{momSC}
\end{eqnarray}
Thus we get the Hamiltonian

\begin{eqnarray}
H^{\pm}_{sh}&=&p_{sh}^{\pm} \acute{R}_{\pm}-L_{sh}^{\pm} \nonumber \\
&=& \frac{m F_{\pm}}{\sqrt{F_{\pm} - F^{-1}_{\pm} \acute{R}^2_{\pm}}} \pm \frac{m^2}{2R} \label{hamSC}
\end{eqnarray}
We can solve (\ref{hamSC}) for $\acute{R}_{\pm}$,

\begin{equation}
\acute{R}_{\pm}= \frac{F_{\pm}}{H^{\pm}_{sh} \mp \frac{m^2}{2R}} \sqrt{(H^{\pm}_{sh} \mp \frac{m^2}{2R})^2 -m^2 F_{\pm}} \label{velacSC} \\
\end{equation}

By comparing (\ref{eqnofmtn}),(\ref{speed}) and (\ref{velacSC}) we see that equations of motion found by Gladush's action (\ref{actionSC}) are the same as those obtained by Israel's method and the Hamiltonian is

\begin{equation}
H^{\pm}_{sh}=\Delta M \label{energyofshell1}
\end{equation}

\subsubsection{Equations of Motion in Infalling Eddington Finkelstein Coordinates}

Here we derive the equations of motion of the shell in Eddington Finkelstein coordinates. From the laws of coordinate transformation explained in \ref{ch:Eddington Finkelstein Coordinates and The Equation of Motion}, we observe that the gauge field in the effective action changes with the change of coordinates to $U=\{-\frac{m}{2R},F_{\pm}^{-1}\frac{m}{2R},0,0\}$. However we can gauge away the radial part because it is only a function of the radial coordinate. The action (\ref{actioncovariant}) in infalling Eddington Finkelstein coordinates is given by

\begin{eqnarray}
I^{\pm}_{sh}&=&-m \int \big( \sqrt{F_{\pm} - 2\acute{R}_{\pm}} \pm \frac{m}{2R}\big) dT_{\pm} \nonumber \\
\acute{R}_{\pm}&\equiv&\frac{dR}{dT_{\pm}} \label{actionEF}
\end{eqnarray}

We calculate the conjugate momentum and the Hamiltonian which will  be a constant of motion since the Lagrangian does not depend explicitly on time.

\begin{eqnarray}
L_{sh}^{\pm}&=&-m \sqrt{F_{\pm} - 2 \acute{R}_{\pm}} \mp \frac{m^2}{2R} \nonumber \\
p_{sh}^{\pm} &=& \frac{m}{\sqrt{F_{\pm} - 2 \acute{R}_{\pm}}} \nonumber \\
H^{\pm}_{sh}&=&\frac{m(F_{\pm} - \acute{R}_{\pm})}{\sqrt{F_{\pm} - 2 \acute{R}_{\pm}}} \pm \frac{m^2}{2R} \label{lagmomhamEF}
\end{eqnarray}
The last of the equations above reduces to

\begin{equation}
K_{\pm}=\frac{(F_{\pm} - \acute{R}_{\pm})}{\sqrt{F_{\pm}- 2 \acute{R}_{\pm}}} \nonumber
\end{equation}
with $K_\pm$ defined by

\begin{eqnarray}
K_{\pm} \equiv  \frac{1}{m}\big( H^{\pm}_{sh} \mp \frac{m^2}{2R} \big) \label{K}
\end{eqnarray}
We now solve for $\acute{R}_{\pm}$.

\begin{equation}
\acute{R}_{\pm}^2 -2 \acute{R}_{\pm}(F_{\pm}-K_{\pm}^2) +F_{\pm}(F_{\pm}-K_{\pm}^2)=0
\end{equation}
The roots of the equation are given by

\begin{equation}
\frac{dR}{dT_{\pm}} \Big|_{\oslash}= (F_{\pm} - K_{\pm}^2) \oslash K_{\pm} \sqrt{K_{\pm}^2 - F_{\pm}} \label{eomEF}
\end{equation}

Comparing (\ref{eqnofmtnEFspeed}) and(\ref{eomEF})  we see that the equation of motion obtained by varying the action is the same as that obtained by Israel's method. By comparing (\ref{eqnofmtn}) and (\ref{K}) we then get

\begin{eqnarray}
K_{\pm}&=&\kappa_{\pm} \nonumber \\
H^{\pm}_{sh} &=&\Delta M \label{KKappaandHamltn}
\end{eqnarray}

Using  (\ref{lagmomhamEF}),(\ref{K}) and (\ref{KKappaandHamltn}) we get

\begin{eqnarray}
p_{sh,\oslash}^{\pm}&=& m \frac{\kappa_{\pm} \oslash \sqrt{\kappa_{\pm}^2 - F_{\pm}}}{F_{\pm}} \label{EFmom}
\end{eqnarray}

Observe that \emph{the Hamiltonian in terms of coordinates on both sides of the shell agrees, but the momentum differs}. This will have an effect on the tunneling calculations.

\section{What is the transmission coefficient ?} \label{ch:Tunneling}

Having proven that the variation of the action (\ref{actioncovariant}) gives the correct equations of motion in Schwarzschild as well as infalling Eddington Finkelstein coordinates, we will use the conjugate momentum from the action in infalling Eddington Finkelstein coordinates to calculate the quantity $exp( -\int p dr)$ with the integral over the future horizon. It will be shown that the answer differs when calculated in terms of internal and external coordinates.\footnote{Alternative derivations of the result are given in \ref{altdrvfrmactn} to prove the robustness of the result.} 

\subsection{Calculation of $exp \left\{- \;Im(\displaystyle{\int p\; dr})\right \}$} \label{ch:tunncalc}

We want to calculate the quantity $exp(- \; Im(\int p\;dr))$ with the integration over the horizon.  We will work in the infalling Eddington Finkelstein coordinates since those cover the future horizon and we want to consider particles coming out of the same. From  (\ref{EFmom}) and (\ref{eqnofmtn})

\begin{eqnarray}
p^{\pm}_{sh, \oslash} &=& m \frac{(\frac{\Delta M}{m} R \mp \frac{m}{2}) \pm \sqrt{(\frac{\Delta M}{m} R \mp \frac{m}{2})^2 - R^2 F_{\pm}  } }{(R-2M_{\pm})}
\end{eqnarray}

The only imaginary part of $\int pdr$, while integrating over classically forbidden regions sandwiched between classically allowed regions comes as a pole for outgoing shells. The pole comes while integrating over the future horizon. We go under the pole to get a lower probability of emission for more energetic particles. Infalling shells do not face a barrier and they do not give any imaginary action. For outgoing shells we have

\begin{equation}
Im(\int p\; dr)= 4 \pi (M_{\pm} \Delta M \mp \frac{m^2}{4}) \label{actionresidue}
\end{equation}
Thus the exponential squared of this quantity is

\begin{equation}
\Gamma_{\pm}  = exp \left\{-2 \;Im(\int p\; dr)\right \} = e^{-8 \pi ( M_{\pm}  \Delta M  \mp \frac{m^2}{4} )}  \label{tunnprob}
\end{equation}

This result is only valid if the shell's turning point is outside the horizon since otherwise the integration will not be over the horizon. For asymptotically free shells, $m=\Delta M$ and we get

\begin{equation}
\Gamma_{\pm} = e^{-2Im(S_{\pm})}=e^{-8 \pi m(M_{\pm} \mp \frac{m}{4})}  \label{v2tunnprob}
\end{equation}

The massless limit ($m \to 0$) which gives the correct equation of motion for a null shell reduces the above expression to 

\begin{equation}
\Gamma_{\pm, massless} = e^{-8 \pi  M_{\pm}  \Delta M }  \label{tunnprobmassless}
\end{equation}

An alternate derivation of these results which brings out the result for the massless case directly instead of by limits is given in \ref{altdrvfrmactn}. 

The interesting point is that the quantity that is usually associated with tunneling transmission coefficient is different when calculated in the inside and outside coordinates. Mathematically this should not come as a surprise for a self gravitating shell has Minkowski vacuum inside and thus faces no barrier from inside coordinates. It does however have a horizon in the outside manifold and thus a barrier.

\subsection{Calculation of $exp \left\{-  \;Im(\displaystyle{\oint p\; dr})\right \}$}

We had argued in section \ref{ch:CanTrans} that the correct expression for tunneling should be $exp \left\{-  \;Im(\oint p\; dr)\right \}$ over the forbidden region. However since the infalling shells face no barrier this expression evaluates to the square root of (\ref{v2tunnprob}) (or  (\ref{tunnprobmassless}) for null shells) and is still different for the inside and the outside manifolds

\subsection{Which coordinates should one trust} \label{ch:trust}

We saw that we have different answers for tunneling transmission coefficient in manifold internal and external to the shell. It looks surprising that even though we get the correct equations of motion from both the actions transmission coefficients turn out to be different. Which is the one we should trust.

First of all the horizons for the two manifolds are at different radial positions. We can only say that tunneling is happened when this information can be conveyed to an asymptotic Minkowskian observer. To that effect the shell should cross the outer manifolds horizon and it makes sense to take that answer seriously. Here those values are given again

\be
e^{-2 Im \int p dr} = e^{-8 \pi m(M_{+} - \frac{m}{4})} 
\ee
and 
\be
e^{- Im \oint p dr} = e^{-4 \pi m(M_{+} - \frac{m}{4})} 
\ee

\section{Conclusion} \label{ch: Conclusion}

We have seen in section \ref{ch:CanTrans} that $\Gamma = exp \left\{-2 \;Im(\int p\; dr)\right \}$ is not invariant under general canonical transformations. We saw that this expression was a simplification of  $exp \left\{- \;Im(\oint p\; dr)\right \}$ which is invariant under canonical transformations. However since the infalling shells face no barrier the latter expression is the square root of the former.

After discussing Israel's equations of motion in section \ref{ch:MtnThShell} and arriving at the same from Gladush's two equivalent actions in section \ref{ch: Reduced Gravity Action For a Thin Shell} we calculated $\Gamma$ for both of the actions in section \ref{ch:Tunneling}. We found that $\Gamma$ was different in internal and external manifolds. However we argued that since we are interested in shells tunneling out of the outer manifold's horizon as our asymptotic observers are outside the shell we gave the expressions for tunneling transmission coefficient in section \ref{ch:trust}.

If one were to assume that the correct expression was a geometric mean of the expressions for inside and outside manifolds one would get $\Gamma = exp (-8\pi m M_{av})$ which is Hawking's result for test shells. This would suggest that there was no self interaction correction. Additionally if we we take the correct expression for $\Gamma = exp \left\{- \;Im(\oint p\; dr)\right \}$ we infact get the square root of Hawking's result. For massless shells we drop the second order terms in $m$  in the above and we see that the answers still differ from the result of \cite{Parikh:1999mf}.

In the authors opinion the correct expression for tunneling transmission has to be invariant under canonical transformations. This suggests that we use the expression $exp(- Im \oint p dr)$. However the temperature we get from this expression is half that of Hawking radiation temperature. The fact that the simple minded generalization of tunneling gives half the correct temperature means that more works needs to be done to understand how tunneling takes place in a background of black holes.

\section*{Acknowledgments}

I would like to thank Valentin D. Gladush, Samir D. Mathur,  Maulik Parikh and Yogesh K. Srivastava for helpful discussions. I would also like to thank Stefano Giusto, Frederick G. Kuehn, Niharika Ranjan, Jason M. Slaunwhite and Yogesh K. Srivastava for their help in correcting the errors in the manuscript.

\newpage

\appendix

\section{Equations of Motion from Israel's Junction Conditions} \label{ch:Israel's Junction Conditions}

The results of Israel \cite{Israel:1966rt} were derived in a very elegant way by Poisson \cite{Poisson:2004rt}. We will extend those results. The results were also extended to Reissner Nordstrom geometries\cite{Heusler:1990in} and we will incorporate those too. We begin with a spherical shell with surface stress energy tensor

\begin{equation}
S_{ab}=\sigma u_a u_b, \quad \sigma=constant 
\end{equation}
dividing space time into two Schwarzschild regions\footnote{By Birkhoff's theorem the geometry inside and oustide will be Reissner Nordstrom} $V_{\pm}$, with coordinates $(t_+,r,\theta,\phi)$ and $(t_-,r,\theta,\phi)$ and with metrics 

\begin{eqnarray}
ds_{\pm}^2&=&-f_{\pm}dt_{\pm}^2+f_{\pm}^{-1}dr^2 + r^2d\Omega^2 \nonumber \\
f_{\pm}&=&1-\frac{2M_{\pm}}{r}+\frac{Q_\pm^2}{r^2}
\end{eqnarray}
When the shell is at a radial coordinate R, the metric of the shell, due to spherical symmetry can be taken as

\begin{equation}
ds_{\Sigma}^2\equiv -d\tau^2 + R^2 d\Omega^2 \label{shellmetric}
\end{equation}
The metric induced on the shell from both sides has to be same by Israel's first junction condition \cite{Israel:1966rt},\cite{Poisson:2004rt}  and we can set it equal to the above.

\begin{eqnarray}
ds_{\Sigma\pm}^2&=&-(F_{\pm}\dot{T}_{\pm}^2-F_{\pm}^{-1}\dot{R}^2)d\tau^2 + R^2 d\Omega^2 \nonumber \\
R&=&R(\tau) \quad T_{\pm} = T_{\pm}(\tau) \quad F_{\pm}=1-\frac{2M_{\pm}}{R}+\frac{Q_\pm^2}{R^2} \nonumber\\
\dot{R}&=&\frac{dR}{d\tau} \quad \dot{T}_{\pm} = \frac{dT_{\pm}}{d\tau}
\end{eqnarray}
Here the shell is at $R(\tau)$ at the time $T_{\pm}(\tau)$ in the regions $V_{\pm}$.
The requirement of the induced metric being continuous becomes

\begin{eqnarray}
F_{\pm} \dot{T}_{\pm}&=&\sqrt{\dot{R}^2+F_{\pm}}  \equiv \kappa_{\pm}(R,\dot{R}) \label{defbeta}
\end{eqnarray}
The velocity of the shell particles is

\begin{eqnarray}
u_{\pm}^{\alpha}&=&\frac{dx_{\pm}^{\alpha}}{d\tau} \nonumber \\
&=&(\dot{T}_{\pm},\dot{R},0,0)
\end{eqnarray}
The normal to the hypersurface formed by the word volume of the shell is gotten by the requirement

\begin{eqnarray}
n_{\pm \,\alpha}n^{\alpha}_{\pm}=1 \quad n_{\pm \alpha}u_{\pm}^{\alpha}=0  \label{normalrequirement}
\end{eqnarray}
Thus

\begin{equation}
n_{\pm \,\alpha}=(-\dot{R},\dot{T}_{\pm},0,0)
\end{equation}
The extrinsic curvature on either side of the shell are defined by

\begin{equation}
K_{ab}\equiv n_{\alpha;\kappa}\frac{dx^{\alpha}}{dy^a}\frac{dx^{\kappa}}{dy^b}
\end{equation}
where $\{x^{\alpha}\}$ are coordinates on the $V_{\pm}$ and $\{y^a\}$ are those on the shell.
The angular components of $K$ are

\begin{eqnarray}
K_{\theta \theta} &=& n_{\theta;\theta} = - \Gamma^{R}_{\theta\theta} n_{R} =\kappa R \nonumber \\
K_{\phi \phi} &=& n_{\phi;\phi} = - \Gamma^{R}_{\phi \phi} n_{R} = \kappa R sin^2(\theta) \label{extcurvang}
\end{eqnarray}
Now we calculate the time component of K

\begin{equation}
K_{\tau\tau}=n_{\alpha;\kappa}u^{\alpha}u^{\kappa}=-a^{\alpha}n_{\alpha}
\end{equation}
on account of (\ref{normalrequirement}).
We have acceleration perpendicular to velocity ($u_{\alpha}u^{\alpha}=constant$ implies $u_{\alpha}a^{\alpha}=0$)

\begin{equation}
-F a^T \dot{T} + F^{-1} a^R \dot{R}=0 \Rightarrow a^T = \frac{a^R\dot{R}}{F^2 \dot{T}} \label{accperpvel}
\end{equation}
Thus
\begin{eqnarray}
K_{\tau\tau}&=&-n_{\alpha}a^{\alpha}= \dot{R} a^{T} - \dot{T} a^{R} \nonumber \\
&=&-a^R \Big[ \dot{T} - \frac{\dot{R}^2}{F^2 \dot{T}}\Big] \nonumber \\
&=&-a^R \frac{\kappa^2 - \dot{R}^2}{\kappa F} \nonumber \\
&=&-\frac{a^R}{\kappa} \label{extcurtemp}
\end{eqnarray}
We find the acceleration in terms of other quantities

\begin{eqnarray}
a^R &=& \frac{d^2 R}{d\tau^2} + \Gamma^R_{TT} \dot{T}^2 + \Gamma^R_{RR} \dot{R}^2 \nonumber \\
&=&\frac{d^2 R}{d\tau^2}  + \frac{1}{2} \Big[ F \partial_R F \dot{T}^2 + F \partial_R F^{-1} \dot{R}^2\Big] \nonumber \\
&=&\frac{d^2 R}{d\tau^2} + \frac{1}{2} \partial_R F \Big[  F \dot{T}^2 - \frac{\dot{R}^2}{F} \Big] \nonumber \\
&=&\frac{d^2 R}{d\tau^2} + \frac{1}{2} \partial_R F \label{accradial}
\end{eqnarray}
Also observe,

\begin{eqnarray}
\frac{\dot{\kappa}}{\dot{R}} = \frac{1}{\dot{R}}  \dot{R} \frac{ \frac{d^2 R}{d\tau^2} + \frac{1}{2} \partial_R F}{\kappa} = \frac{a^R}{\kappa} \label{seemunr}
\end{eqnarray}
Thus from (\ref{extcurvang}),(\ref{accradial}), (\ref{seemunr}) and (\ref{shellmetric}) we get the extrinsic curvatures to be

\begin{eqnarray}
K^{\tau}_{\pm\tau}&=&\frac{\dot{\kappa}_{\pm}}{\dot{R}}\nonumber\\
K^{\theta}_{\pm\theta}&=&K^{\phi}_{\pm}{\phi}=\frac{\kappa_{\pm}}{R}
\end{eqnarray}
From the second junction condition \cite{Israel:1966rt},\cite{Poisson:2004rt} 

\begin{equation}
S^a_{\phantom{a}b}=-\frac{1}{8\pi}\big( [K^a_{\phantom{a}b}]-[K]\delta^a_{\phantom{a}b}\big)
\end{equation}
where $K\equiv K_{ab}h^{ab}$ and  $[A] \equiv A_+ - A_-$. \newline
Thus

\begin{eqnarray}
K&=&\frac{\dot{\kappa}}{\dot{R}} + 2 \frac{\kappa}{R} \nonumber \\
S^{\tau}_{\tau}&=&\frac{[\kappa]}{4\pi R}  = - \sigma \nonumber \\
S^{\theta}_{\theta}&=&S^{\phi}_{\phi}=\frac{[\kappa]}{8 \pi R} + \frac{[\dot{\kappa}]}{\dot{R}}=0
\end{eqnarray}
The solution to the second of these is

\begin{equation}
[\kappa]R=constant
\end{equation}
And using this in the first

\begin{equation}
[\kappa]R = - \sigma 4 \pi R^2 = -constant  \equiv -m
\end{equation}
Solving for $\dot{R}$ using (\ref{defbeta}) we get two versions which are equivalent 

\begin{eqnarray}
\dot{R}^2 &=& \frac{1}{m^2} \big[ \Delta M - \frac{\Delta Q^2  \pm m^2}{2R}\big]^2 -F_{\pm} 
\end{eqnarray}
Where $\Delta M= M_+ - M_-$ and $\Delta Q^2 = Q_+^2 - Q_-^2$. We can also get

\begin{eqnarray}
\kappa_{\pm}=\frac{1}{m}\big[ \Delta M  - \frac{\Delta Q^2  \pm m^2}{2R}\big]
\end{eqnarray}
Thus the complete solution is

\begin{eqnarray}
\kappa_{\pm}&=&\frac{1}{m}\big[ \Delta M  - \frac{\Delta Q^2  \pm m^2}{2R}\big] \nonumber \\
\dot{T}_{\pm}&=& \frac{\kappa_{\pm}}{F_{\pm}} \nonumber \\
\dot{R}^2 &=& \kappa_{\pm}^2- F_{\pm} \label{eqnofmtnappendix}
\end{eqnarray}

\section{Eddington Finkelstein Coordinates and the Equations of Motion} \label{ch:Eddington Finkelstein Coordinates and The Equation of Motion}

The transformation law between the Schwarzschild coordinates $(t_{sc},r_{sc},\theta,\phi)$ and Infalling Eddington Finkelstein coordinates $(t_{ef},r_{ef},\theta,\phi)$ is

\begin{eqnarray}
dr_{ef}&=&dr_{sc} \nonumber \\
dt_{ef} &=& dt_{sc} + f^{-1} dr_{sc}
\end{eqnarray}
So the radial coordinate is the same. The metric in these coordinates is

\begin{eqnarray}
ds_{\pm}^2 = -f_{\pm} dt_{ef,\pm}^2 + 2 dt_{ef,\pm} dr + r^2 d\Omega^2
\end{eqnarray}
With this and the equations of motion in Schwarzschild coordinates (\ref{eqnofmtnappendix}) we have in Infalling Eddington Finkelstein coordinates (with $\oslash=\pm$ for outgoing and infalling shells)

\begin{eqnarray}
\dot{T}_{ef,\oslash,\pm}&=&\dot{T}_{sc,\pm} \oslash F_{\pm}^{-1} \dot{R}_{sc} \nonumber \\
\dot{R}_{ef}&=&\dot{R}_{sc}
\end{eqnarray}
Explicitely

\begin{eqnarray}
\kappa_{\pm}&=&\frac{1}{m}\big[ \Delta M \mp \frac{m^2}{2R}\big] \nonumber \\
\dot{T}_{ef,\oslash,\pm}&=&\frac{\kappa_{\pm} \oslash  \dot{R}_{ef}}{F_{\pm}} \nonumber \\
\dot{R}_{ef}&=&\kappa_{\pm}^2 - F_{\pm}
\end{eqnarray}

\section{Alternative Derivations of $exp \left\{- 2\;Im(\displaystyle{\int p\; dr})\right \}$} \label{altdrvfrmactn}

In section \ref{ch:tunncalc} we had evaluated the value of $ exp \left\{- \;Im(\int p\; dr)\right \}$ and here we will re-derive the same by a) The Hamilton-Jacobi equation (as imaginary part of the action) and b) by using a result of black-hole emission of charged particles due to Hawking and Hartle \cite{Hartle:1976tp}.

\subsection{Re-derivation using Hamilton-Jacobi Equation}

This section is based on obtaining the action from the Hamilton-Jacobi equation. This method was used in \cite{Srinivasan:1998ty} to arrive at the usual expression for Hawking radiation without  self-interaction correction.
\newline
\newline
{\bf Massive Case}
\newline
The covariant Lagrangian (\ref{actioncovariant}), momenta and eqn. of motion are

\begin{eqnarray}
\mathfrak{L}^{\pm}&=&-m\sqrt{-\dot{x^{\mu}}\dot{x_{\mu}}} \pm U_{\mu}\dot{x}^{\mu} \nonumber \\
\mathfrak{p}^{\pm}_{\mu}&=& m\dot{x}_{\mu} \pm U_{\mu} \nonumber \\
(\mathfrak{p}^{\pm} \mp U)^2 + m^2&=&0  
\end{eqnarray}
where  $U=(-\frac{m^2}{2R},0)$.

We now apply the Hamilton-Jacobi method.  Observe that $\mathfrak{L}^{\pm}$ is independent of $T$ so we can choose the following ansatz for the action (taking the total energy the shell as from (\ref{energyofshell1}) and (\ref{KKappaandHamltn}) as $\Delta M$).

\begin{equation}
S_{\oslash,\pm}=-\Delta M T_{\pm} + W_{\oslash,\pm}(R)
\end{equation}
Replacing $\mathfrak{p}_{\mu}$ in the equation of motion by $\partial_{\mu} S$ we get

\begin{eqnarray}
-F_{\pm}(W'_{\oslash,\pm})^2+2(-\Delta M \pm \frac{m^2}{2R})W'_{\oslash,\pm} +m^2&=&0 \nonumber \\
W'_{\oslash,\pm}=\frac{(\Delta MR \mp \frac{m^2}{2}) \pm \sqrt{(\Delta MR \mp \frac{m^2}{2})^2-m^2 R^2 F_{\pm}}}{R-2M_{\pm}}
\end{eqnarray} 

The imaginary part of the action is the same as imaginary part of $W(R)$ which is gotten by going under the pole. Hence for outgoing particles we recover

\begin{equation}
\Gamma_{\pm}  = e^{-2Im(S_{\pm})} = e^{-8 \pi ( M_{\pm}  \Delta M  \mp \frac{m^2}{4} )} \label{covtunn}
\end{equation}
\newline
\newline
\newline
{\bf Massless Case}
\newline
The following Hamilton Jacobi ansatz can be used for the action
  
\begin{equation}
S_{\oslash,\pm}=-ET_{\pm} + W_{\oslash,\pm}(R)
\end{equation}
The light like shells equation of motion can be obtained by taking the massless limit of the momentum equation of massive particles

\begin{equation}
p_{\oslash,\pm}^2=0
\end{equation}
Thus in Eddington Finkelstein coordinates we get

\begin{equation}
-F_{\pm}W'(R)_{\oslash,\pm}^2+2EW'(R)_{\oslash,\pm}=0
\end{equation}
The solution for the outgoing shell is

\begin{equation}
W(R)_{\pm}=\int dR \frac{2ER}{R-2M_{\pm}}
\end{equation}
The imaginary part is gotten by shifting the contour under the pole giving

\begin{equation}
\Gamma_{\pm} = e^{-2Im(S_{\pm})} =e^{-8\pi E M_{\pm}} \label{tunnmassless}
\end{equation}
This matches the expression gotten by the massless limit (\ref{tunnprobmassless}).

\subsection{Re-derivation using Hawking's Charged Hole Radiance Formula}

In the paper \cite{Hartle:1976tp} Hawking and Hartle argue that the transmission coefficient of emission of a charged particle from a charged hole is given by

\begin{eqnarray}
P_{emission}=e^{-4 \pi  (E-\frac{qQ}{R_H})/f'(R_H)} P_{absorption}
\end{eqnarray}
where q is the mass of emitted particle and Q of the RN hole and  $R_H$ is the outer horizon \footnote{Here outer horizon  means the outer horizon of a multiple horizon black hole and not the horizon of the outer manifold of the shell. In \cite{Hartle:1976tp} since there are only test shells, there is just one manifold.}.

   Although the emission transmission coefficient for self energy was not given in the paper it is easy to extend the result without going into the details of the calculation. This is so because the self interaction comes as a U(1) gauge potential and the electromagnetic interaction is also a U(1) gauge interaction. So we can get the desired result by the following replacement
   
\begin{equation}
\frac{qQ}{R} \to \pm \frac{m^2}{2R} \label{replacement} \quad \text{for $V^{\pm}$}
\end{equation}

 By this replacement we get the result for emission of a shell with energy $\Delta M$ as (understand here $R_H$ would become the the horizon of whichever manifold we are considering, $R_H=2M_{\pm}$)

\begin{eqnarray}
P_{emission}^{\pm}=e^{-8 \pi M_{\pm} (\Delta M \mp \frac{m^2}{4M_{\pm}})} P_{absorption}^{\pm} \label{emmsprobhawk}
\end{eqnarray}
Thus extending Hawking and Hartles method also gives the same result as (\ref{tunnprob}) and (\ref{covtunn}).

\section{Meeting Newton's Laws: Motivating the Action} \label{ch:Meeting Newton's Laws}

The derivation of equations of motion is self consistent. We did not have anything as an external force on the system. The movement is 'of its own' if you will. The question is can we have any notion of Newton's second law so that we can attribute some kind of force on the shell. Let's see what happens to the acceleration. On account of (\ref{seemunr}) and (\ref{accperpvel}) we have

\begin{eqnarray}
a^{R}&=& \frac{\kappa_{\pm} \dot{\kappa}_{\pm}}{\dot{R}} = F_{\pm} \frac{\dot{\kappa}_{\pm}}{\dot{R}} \dot{T}_{\pm}=g^{RR}\frac{\dot{\kappa}_{\pm}}{\dot{R}} \dot{T}_{\pm} \label{accrad} \\
a^{T}&=&\frac{\dot{\kappa}_{\pm}}{F_{\pm}} = -\frac{1}{F_{\pm}} \Big (-\frac{\dot{\kappa}_{\pm}}{\dot{R}} \Big )\dot{R} = - g^{TT}\frac{\dot{\kappa}_{\pm}}{\dot{R}} \dot{R} \label{acctemp}
\end{eqnarray}
we can immediately see that (\ref{accrad}) and (\ref{acctemp}) are of the form

\begin{equation}
a^{\mu}_{\pm} = G^{\mu\nu}_{\pm}u_{\pm \; \nu}
\end{equation}
if we identify

\begin{equation}
G_{\pm \; RT} =  \frac{\dot{\kappa}_{\pm}}{\dot{R}}
\end{equation}
If we have a $\kappa$ of the form 

\begin{equation}
\kappa_{\pm} = (\eta_{\pm} - \frac{\gamma_{\pm}}{R})
\end{equation}
for some constants $\eta_\pm$ and $\gamma_\pm$ we get

\begin{equation}
G_{\pm \;RT} =\frac{\gamma_{\pm}}{R^2}
\end{equation}
And we see we can then explain this motion by a gauge potential $A$ with $G=dA$ with

\begin{equation}
A_{\pm\;T}=-\frac{\gamma_{\pm}}{R}
\end{equation}
This result matches with that in \cite{Gladush:2000rs}.


\newpage

\begin{thebibliography}{99}

\bibitem{Hawking:1974sw}
  S.~W.~Hawking,
  Commun.\ Math.\ Phys.\  {\bf 43}, 199 (1975)
  [Erratum-ibid.\  {\bf 46}, 206 (1976)].

\bibitem{Hartle:1976tp}
  J.~B.~Hartle and S.~W.~Hawking,
  Phys.\ Rev.\ D {13}, 2188 (1976).
  
\bibitem{Giddings:1995gd}
  S.~B.~Giddings,
   ``The Black hole information paradox,''
  %
  arXiv:hep-th/9508151.

\bibitem{Israel:1966rt}
  W.~Israel,
  Nuovo Cim.\ B {44S10}, 1 (1966)
  [Erratum-ibid.\ B {48}, 463 (1967\ NUCIA,B44,1.1966)].

\bibitem{Kraus:1994by}
  P.~Kraus and F.~Wilczek,
  Nucl.\ Phys.\ B {433}, 403 (1995)
  [arXiv:gr-qc/9408003].

\bibitem{Parikh:1999mf}
  M.~K.~Parikh and F.~Wilczek,
  Phys.\ Rev.\ Lett.\  {85}, 5042 (2000)
  [arXiv:hep-th/9907001].
  

  
 
  

  
\bibitem{Akhmedov:2006un}
  E.~T.~Akhmedov, V.~Akhmedova, T.~Pilling and D.~Singleton,
  arXiv:hep-th/0605137.
  
\bibitem{Wu:2006nj}
  S.~Q.~Wu and Q.~Q.~Jiang,
  arXiv:hep-th/0603082.
  
\bibitem{Hu:2006ek}
  Y.~p.~Hu, J.~y.~Zhang and Z.~Zhao,
  arXiv:gr-qc/0601018.
  
\bibitem{Arzano:2005jt}
  M.~Arzano,
  Phys.\ Lett.\ B {\bf 634}, 536 (2006)
  [arXiv:gr-qc/0512071].
  
\bibitem{Radinschi:2005ap}
  I.~Radinschi,
  arXiv:gr-qc/0511142.
  
\bibitem{Nadalini:2005xp}
  M.~Nadalini, L.~Vanzo and S.~Zerbini,
  J.\ Phys.\ A {\bf 39}, 6601 (2006)
  [arXiv:hep-th/0511250].
  
\bibitem{Medved:2005yf}
  A.~J.~M.~Medved and E.~C.~Vagenas,
   ``On Hawking radiation as tunneling with back-reaction,''
  %
  Mod.\ Phys.\ Lett.\ A {\bf 20}, 2449 (2005)
  [arXiv:gr-qc/0504113].
  
\bibitem{Kerner:2006vu}
  R.~Kerner and R.~B.~Mann,
   ``Tunnelling, temperature and Taub-NUT black holes,''
  %
  Phys.\ Rev.\ D {\bf 73}, 104010 (2006)
  [arXiv:gr-qc/0603019].
  
 
  
\bibitem{Melnikov:2002qd}
  K.~Melnikov and M.~Weinstein,
  Int.\ J.\ Mod.\ Phys.\ D {\bf 13}, 1595 (2004)
  [arXiv:hep-th/0205223].
  
\bibitem{Zhang:2005mn}
  J.~g.~Zhang, Y.~p.~Hu and Z.~Zhao,
  arXiv:hep-th/0512121.
  
\bibitem{Einhorn:2005bi}
  M.~B.~Einhorn,
  arXiv:hep-th/0510148.
  
  
\bibitem{Greene:2005wk}
  B.~Greene, M.~Parikh and J.~P.~van der Schaar,
  arXiv:hep-th/0512243.
  
  
\bibitem{Gladush:2000rs}
  V.~D.~Gladush,
  J.\ Math.\ Phys.\  {42}, 2590 (2001)
  [arXiv:gr-qc/0001073].
  
\bibitem{Rubakov:2002fi}
  V.~A.~Rubakov,
  ``Classical theory of gauge fields'', Princeton University Press (2002).

\bibitem{Poisson:2004rt}
  E.~Poisson,
  ``A Relativist's Toolkit'', Cambridge University Press (2004).
  
\bibitem{Srinivasan:1998ty}
  K.~Srinivasan and T.~Padmanabhan,
   ``Particle production and complex path analysis,''
  %
  Phys.\ Rev.\ D {\bf 60}, 024007 (1999)
  [arXiv:gr-qc/9812028].
  
 


 
\bibitem{Heusler:1990in}
  M.~Heusler, C.~Kiefer and N.~Straumann,
  Phys.\ Rev.\ D {42}, 4254 (1990)
  [Erratum-ibid.\ D {44}, 1342 (1991)].
 
 \end{thebibliography}
\end{document}